\documentclass[aps,pre,showkeys,
longbibliography,
reprint,
twocolumn,
nofootinbib,
floatfix]{revtex4-2}

\usepackage[utf8]{inputenc}
\usepackage{amsmath}
\usepackage{amsfonts}
\usepackage{amssymb}
\usepackage{wasysym}
\usepackage{graphicx}
\usepackage{dcolumn}
\usepackage{tikz}

\usepackage[labelformat=simple]{subcaption}

\usepackage{listings}
\definecolor{codegreen}{rgb}{0,0.6,0}
\definecolor{codegray}{rgb}{0.5,0.5,0.5}
\definecolor{codepurple}{rgb}{0.58,0,0.82}
\definecolor{backcolour}{rgb}{0.95,0.95,0.92}
\lstdefinestyle{mystyle}{
    backgroundcolor=\color{backcolour},   
    commentstyle=\color{codegreen},
    keywordstyle=\color{magenta},
    numberstyle=\tiny\color{codegray},
    stringstyle=\color{codepurple},
    basicstyle=\ttfamily\footnotesize,
    breakatwhitespace=false,         
    breaklines=true,                 
    captionpos=t,                    
    keepspaces=true,                 
    numbers=left,                    
    numbersep=5pt,                  
    showspaces=false,                
    showstringspaces=false,
    showtabs=false,                  
    tabsize=2
}
\usepackage[colorlinks=true,hyperfootnotes=true,breaklinks=true]{hyperref}
\usepackage[capitalise]{cleveref}

\begin{document}
\title{Universality of random-site percolation thresholds for two-dimensional complex non-compact neighborhoods}
\author{Krzysztof Malarz}
\thanks{\includegraphics[width=10pt]{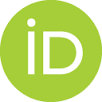}~\href{https://orcid.org/0000-0001-9980-0363}{0000-0001-9980-0363}}
\email{malarz@agh.edu.pl}
\affiliation{AGH University, Faculty of Physics and Applied Computer Science, al. Mickiewicza 30, 30-059 Krak\'ow, Poland}

%% #############################################################
\begin{abstract}
The phenomenon of percolation is one of the core topics in statistical mechanics. 
It allows one to study the phase transition known in real physical systems only in a purely geometrical way.
In this paper, we determine thresholds $p_c$ for random site percolation in triangular and honeycomb lattices for all available neighborhoods containing sites from the sixth coordination zone.
The results obtained (together with the percolation thresholds gathered from the literature also for other complex neighborhoods and also for a square lattice) show the power-law dependence $p_c\propto(\zeta/K)^{-\gamma}$ with
$\gamma=0.526(11)$, $0.5439(63)$ and $0.5932(47)$,
for honeycomb, square, and triangular lattice, respectively, and $p_c\propto\zeta^{-\gamma}$ with
$\gamma=0.5546(67)$
independently on the underlying lattice. 
The index $\zeta=\sum_i z_i r_i$ stands for an average coordination number weighted by distance, that is, depending on the coordination zone number $i$, the neighborhood coordination number $z_i$ and the distance $r_i$ to sites in $i$-th coordination zone from the central site.
The number $K$ indicates lattice connectivity, that is, $K=3$, 4 and 6 for the honeycomb, square and triangular lattice, respectively.
\end{abstract}
%% #############################################################


\date{\today}

\maketitle

%% #############################################################
\section{Introduction}
%% #############################################################

The percolation \cite{bookDS,bookBB,bookMS,bookHK} is one of a core topics in statistical physics as it allows for studying phase transitions and their properties in only geometrical fashion, i.e., without heating or cooling anything (except of paying unconscionable invoices for electricity in the computer centers). 
Although originated from a rheology \cite{Broadbent1957,Hammersley1957} (and still applied there \cite{ISI:000496837300028,*Bolandtaba2011,*Mun2014,*ISI:000524118200031}) the application of percolation theory range from 
forest fires \cite{Kaczanowska2002,*Guisoni2011,*Simeoni2011,*Camelo-Neto2011,*Abades2014}
to disease propagation \cite{2101.00550}, 
not omitting problems originated in hard physics (including magnetic \cite{PhysRevB.97.165121,*ISI:000400959000004,*PhysRevB.94.054407,*Yiu,*Grady_2023} and electric \cite{ISI:000419615800018,*ISI:000462936100013,*ISI:000514848600043} properties of solids)
but also with implications for:
nanoengineering \cite{Xu_2014};
materials chemistry \cite{Alguero_2020,*Meloni2022};
agriculture \cite{ISI:000518460000003,*PhysRevE.109.014304};
sociology \cite{ISI:000168785900005};
terrorism \cite{Galam_2002269,*Galam_2003695};
urbanization \cite{ISI:000523958600016};
dentistry \cite{Beddoe_2023};
information transfer \cite{Cirigliano_2023};
computer networks \cite{Liu20231862};
psychology of motivation \cite{2206.14226};
and finances \cite{Bartolucci2020}
(see References~\onlinecite{Saberi2015,Li_2021,Sahimi2023517} for the most recent reviews also on fractal networks \cite{fractalfract7030231} or explosive percolation \cite{Cho2022,*Li2023,*Wu2022,*Luo2023,*Hagiwara2022}).

The phase transition mentioned above is first of all characterized by a critical parameter called \emph{percolation threshold} $p_c$ and much effort went into searching for a universal formula that allows for the prediction $p_c$ based solely on the scalar characteristics of a lattice or a network topology, where the percolation phenomenon occurs.
Probably, searching for such dependencies is not different much from searching for the alchemic formula for the philosopher's stone---allowing for converting anything (or at least something) into gold. 
Anyway, such attempts of proposing universal formula for percolation threshold were more or less successfully made earlier.

For instance, \citeauthor{PhysRevE.53.2177} \cite{PhysRevE.53.2177} proposed an universal formula 
\begin{subequations}
\label{eq:Galam-Mauger-PRE.53.2177}
%% --------------------------------------------------------------
\begin{equation} \label{eq:Galam-Mauger} 
p_c= \dfrac{p_0}{{[(d-1)(z-1)]}^a}
\end{equation}
%% --------------------------------------------------------------
depending on the connectivity of the lattice $z$ and its dimension $d$.
For a site percolation problem they identified two groups of lattices, i.e., two sets of parameters $p_0$ and $a$.
Their paper was immediately criticized by \citeauthor{ISI:A1997WD54600080} \cite{ISI:A1997WD54600080,PhysRevE.55.1230} who indicated two lattices with identical $z$ and $d$ but different values of $p_c$ associated with these lattices.
For two-dimensional lattices the Galam--Mauger formula reduces to
%% --------------------------------------------------------------
\begin{equation} \label{eq:Galam-Mauger-2d} 
p_c= \dfrac{p_0}{(z-1)^a},
\end{equation}
%% --------------------------------------------------------------
\end{subequations}
with $p_0=0.8889$ and $a=0.3601$ for triangular, square and honeycomb lattices \cite{PhysRevE.53.2177}. Their studies were extended to anisotropic lattices without equivalent nearest neighbors, non-Bravais lattices with two atom unit cells, and quasicrystals which required the substitution of $z$ in Equation~\eqref{eq:Galam-Mauger-PRE.53.2177} by an effective (non integer) value $z_\text{eff}$ \cite{Galam_1997322,PhysRevE.59.1278}.

Very recently, \citeauthor{PhysRevE.105.024105} \cite{PhysRevE.105.024105} in extensive numerical simulations showed that all Archimedean lattices (uniform tilings, i.e., lattices built of repeatably sequences of tails of regular polygons able to cover a two-dimensional plane) exhibit a simple relation
%% ==============================================================
\begin{subequations}
\label{eq:Ziff}
%% --------------------------------------------------------------
\begin{equation} \label{eq:Ziff-a}
p_c=c_1/z,
\end{equation}
%% -------------------------------------------------------------
which due to finite size effects should be modified by constant term $b$
%% --------------------------------------------------------------
\begin{equation} \label{eq:Ziff-b}
p_c=c_2/z-b.
\end{equation}
%% --------------------------------------------------------------
\end{subequations}
%% ==============================================================
For example, for the square lattice and extended compact neighborhoods, these constants are $c_2=4.527$ and $b=3.341$ \cite{PhysRevE.103.022126}.
In two dimensions, for Archimedean lattices up to the 10-th coordination zone \cite{PhysRevE.105.024105}, correlations are also seen by plotting 
\begin{equation}
\label{PhysRevE.105.024105}
z \text{ versus } -1/\ln(1-p_c).
\end{equation}
Yet another investigated by \citeauthor{Galam_Mauger_1994a} \cite{Galam_Mauger_1994a,Galam_Mauger_1994b} formulas included 
\begin{equation}
\label{eq:Galam-Mauger-sqrt}
p_c = 1/\sqrt{z-1} 
\end{equation}
or by \citeauthor{Koza_2014} \cite{Koza_2014,Koza_2016} 
\begin{equation}
\label{eq:Koza}
p_c=1-\exp(d/z).
\end{equation}

The formula \eqref{eq:Ziff-a} works well also for distorted lattices \cite{PhysRevE.99.012117,PhysRevE.106.034109}, where lattice distortion means random moving of lattice nodes not too far from their regular position in non-distorted lattices. 
In this case, the number of sites in the neighborhood $z$ should be replaced by an average site degree $\bar z$ \cite{2306.05513}.

The studies mentioned above were concentrated in compact neighborhoods.
When holes in the neighborhoods are taken into account, there is a strong degeneration of $p_c$ on total $z$, and \Cref{eq:Galam-Mauger-PRE.53.2177,PhysRevE.105.024105,eq:Galam-Mauger-sqrt,eq:Ziff,eq:Koza}---which depend solely on the lattice dimension $d$ and connectivity $z$---must fail.
To avoid this $p_c(z)$ degeneracy in the case of triangular lattice, a weighted square distance
%% --------------------------------------------------------------
\begin{equation} \label{eq:xi} 
\xi=\sum_i r_i^2 z_i/i
\end{equation}
%% --------------------------------------------------------------
was proposed, where $z_i$ is the number of sites in the given neighbourhood in $i$-th coordination zone and these sites distance to the central site in neighbourhood is $r_i$ \cite{2102.10066}.
Unfortunately, the clear dependence 
%% --------------------------------------------------------------
\begin{equation} \label{eq:pcvsxi} 
p_c\propto \xi^{-\gamma}  
\end{equation}
%% ---------------------------------------------------------------
(with $\gamma^\xi_{\textsc{tr}}\approx 0.710(19)$) is lost for the honeycomb lattice \cite{2204.12593}.
Thus, instead, the weighted coordination number
%% ---------------------------------------------------------------
\begin{equation} \label{eq:zeta} 
\zeta=\sum_i z_i r_i
\end{equation}
%% ---------------------------------------------------------------
was proposed \cite{2204.12593} which gives a nice power law
%% ---------------------------------------------------------------
\begin{equation} \label{eq:pcvszeta} 
p_c\propto\zeta^{-\gamma}
\end{equation}
%% ---------------------------------------------------------------
with $\gamma^\zeta_{\textsc{hc}}\approx 0.4981(90)$.
As $\gamma^\zeta_{\textsc{hc}}$ is very close to $\frac 12$ also the dependence 
%% --------------------------------------------------------------
\begin{equation} \label{eq:pcvssqrtzeta} 
p_c=c_3/\sqrt{\zeta}
\end{equation}
%% ---------------------------------------------------------------
was checked yielding $c_3\approx 1.2251(99)$ \cite{2204.12593}.

%% ===============================================================
\begin{figure*}[htbp]
%% ---------------------------------------------------------------
\begin{subfigure}[b]{0.16\textwidth}
\caption{\label{fig:tr-1nn}}
\includegraphics[width=0.99\textwidth]{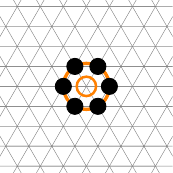}
\end{subfigure}
\hfill %% ---------------------------------------------------------------
\begin{subfigure}[b]{0.16\textwidth}
\caption{\label{fig:tr-2nn}}
\includegraphics[width=0.99\textwidth]{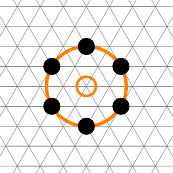}
\end{subfigure}
\hfill %% ---------------------------------------------------------------
\begin{subfigure}[b]{0.16\textwidth}
\caption{\label{fig:tr-3nn}}
\includegraphics[width=0.99\textwidth]{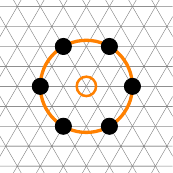}
\end{subfigure}
\hfill %% ---------------------------------------------------------------
\begin{subfigure}[b]{0.16\textwidth}
\caption{\label{fig:tr-4nn}}
\includegraphics[width=0.99\textwidth]{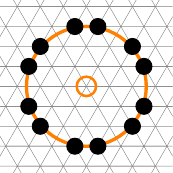}
\end{subfigure}
\hfill %% ---------------------------------------------------------------
\begin{subfigure}[b]{0.16\textwidth}
\caption{\label{fig:tr-5nn}}
\includegraphics[width=0.99\textwidth]{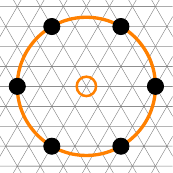}
\end{subfigure}
\hfill %% ---------------------------------------------------------------
\begin{subfigure}[b]{0.16\textwidth}
\caption{\label{fig:tr-6nn}}
\includegraphics[width=0.99\textwidth]{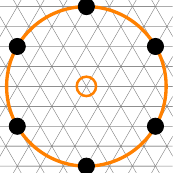}
\end{subfigure}
%% ---------------------------------------------------------------
\caption{\label{fig:tr-neighbors-basics}Basic neighborhoods corresponding to subsequent coordination zones $i=1,\cdots,6$ in the triangular lattice. The symbol $r$ stands for the Euclidean distance of the black sites from the central one, and $z$ indicates the number of sites in the neighborhood.
\subref{fig:tr-1nn} \textsc{tr}-1: $i=1$, $r^2=1$,  $z=6$,
\subref{fig:tr-2nn} \textsc{tr}-2: $i=2$, $r^2=3$,  $z=6$,
\subref{fig:tr-3nn} \textsc{tr}-3: $i=3$, $r^2=4$,  $z=6$,
\subref{fig:tr-4nn} \textsc{tr}-4: $i=4$, $r^2=7$,  $z=12$,
\subref{fig:tr-5nn} \textsc{tr}-5: $i=5$, $r^2=9$,  $z=6$,
\subref{fig:tr-6nn} \textsc{tr}-6: $i=6$, $r^2=12$, $z=6$.}
\end{figure*}
%% ===============================================================

%% ===============================================================
\begin{figure*}[htbp]
%% ---------------------------------------------------------------
\begin{subfigure}[b]{0.16\textwidth}
\caption{\label{fig:hc-1nn}}
\includegraphics[width=0.99\textwidth]{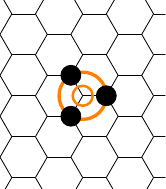}
\end{subfigure}
\hfill %% ---------------------------------------------------------------
\begin{subfigure}[b]{0.16\textwidth}
\caption{\label{fig:hc-2nn}}
\includegraphics[width=0.99\textwidth]{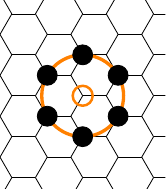}
\end{subfigure}
\hfill %% ---------------------------------------------------------------
\begin{subfigure}[b]{0.16\textwidth}
\caption{\label{fig:hc-3nn}}
\includegraphics[width=0.99\textwidth]{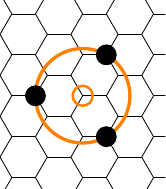}
\end{subfigure}
\hfill %% ---------------------------------------------------------------
\begin{subfigure}[b]{0.16\textwidth}
\caption{\label{fig:hc-4nn}}
\includegraphics[width=0.99\textwidth]{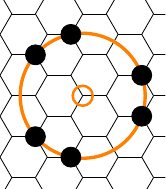}
\end{subfigure}
\hfill %% ---------------------------------------------------------------
\begin{subfigure}[b]{0.16\textwidth}
\caption{\label{fig:hc-5nn}}
\includegraphics[width=0.99\textwidth]{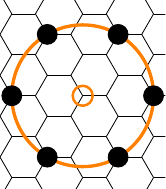}
\end{subfigure}
%% ---------------------------------------------------------------
\begin{subfigure}[b]{0.16\textwidth}
\caption{\label{fig:hc-6nn}}
\includegraphics[width=0.99\textwidth]{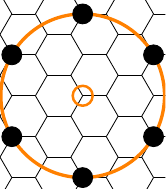}
\end{subfigure}
%% ---------------------------------------------------------------
\caption{\label{fig:hc-neighbors-basics}Basic neighborhoods corresponding to subsequent coordination zones $i=1,\cdots,6$ on the honeycomb lattice.
The symbol $r$ stands for the Euclidean distance of the black sites from the central one, and $z$ indicates the number of sites in the neighborhood.
\subref{fig:hc-1nn} \textsc{hc-1}: $i=1$, $r^2=1$, $z=3$.
The lattices
\subref{fig:hc-2nn} \textsc{hc}-2: $i=2$, $r^2=3$, $z=6$,
\subref{fig:hc-5nn} \textsc{hc}-5: $i=5$, $r^2=9$, $z=6$ and 
\subref{fig:hc-6nn} \textsc{hc}-6: $i=6$, $r^2=12$, $z=6$ are equivalent to a triangular lattice \textsc{tr-1} [\Cref{fig:tr-1nn}] with enlarged lattice constants $\sqrt 3$, $3$ and $2\sqrt 3$ times, respectively.
\subref{fig:hc-4nn} \textsc{hc}-4: $i=4$, $r^2=7$, $z=6$.
The lattice \subref{fig:hc-3nn} \textsc{hc-3} ($i=3$, $r^2=4$, $z=3$,) is equivalent to \textsc{hc-1} [\Cref{fig:hc-1nn}] with a lattice constant twice larger than for \textsc{hc-1}}
\end{figure*}
%% ===============================================================

Very recently, we tested formulas \eqref{eq:pcvsxi} and \eqref{eq:pcvszeta} also for the square lattice up to the sixth coordination zone and found that Eq.~\eqref{eq:pcvszeta} also holds for a square lattice with $\gamma^\zeta_{\textsc{sq}}\approx 0.5454(60)$ \cite{2303.10423}.

Our results show that for all three (square, triangular, and honeycomb) lattice shapes, the power law is recovered in dependence of $p_c(\zeta/K)$, where $K$ is the connectivity of the network with the nearest-neighbour interaction, that is, with $K=3$, 4 and 6 for the honeycomb, square, and triangular lattice, respectively. 
On the other hand, independent of the lattice topology, we see a more or less clear power law $p_c(\zeta)$ for the data obtained on the values of $p_c$ for the three lattices with complex neighbourhoods containing sites up to the sixth coordination zone.

%% #############################################################
\section{Methodology}
%% #############################################################

In this paper---using exactly the same methodology as that used to study percolation in a square lattice with complex neighborhoods that contain sites up to the sixth coordination zone \cite{2303.10423}---we extend our previous studies for sites up to the sixth coordination zone for triangular (\Cref{fig:tr-neighbors-basics}) and honeycomb (\Cref{fig:hc-neighbors-basics}) lattices.
Namely, using the fast Monte Carlo scheme proposed by \citeauthor{NewmanZiff2001} \cite{NewmanZiff2001} and the finite-size scaling theory \cite{Finite-Size_Scaling_Theory_1990,Guide_to_Monte_Carlo_Simulations_2009} we found 64 values of percolation thresholds for complex neighborhoods containing sites from the sixth coordination zone.

In Supplemental Material \cite{SM-EL12139} the mapping of the 6th coordination zone in the honeycomb lattice into the brick-wall-like square lattice (as proposed in Reference~\onlinecite{PhysRevE.60.275}) is presented in Figure 1 
in Appendix A 
together with Listing 1 
(for \textsc{tr-6} neighborhood) and Listing 2 
(for \textsc{hc-6} neighborhood) showing implementations of \verb|boundaries()| functions to be replaced in original Newman--Ziff algorithm \cite{NewmanZiff2001}.
The mapping of the 1st to 5th coordination zones in the honeycomb lattice into the brick-wall-like square lattice are presented in Figure 3 in Reference~\cite{2204.12593}.

%% #############################################################
\section{Results}
%% #############################################################

%% =============================================================
\begin{figure}[htbp]
%% --------------------------------------------------------------
\begin{subfigure}[b]{0.4\textwidth}
\caption{\label{fig:Smax-tr-123456}} 
\includegraphics[width=0.91\textwidth]{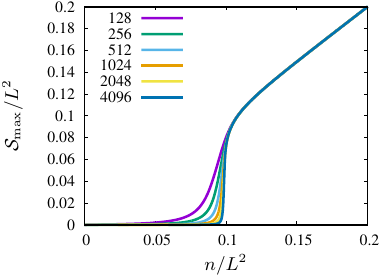}
\end{subfigure}
%% --------------------------------------------------------------
\begin{subfigure}[b]{0.4\textwidth}
\caption{\label{fig:PmaxLbetanu-tr-123456}} 
\includegraphics[width=0.91\textwidth]{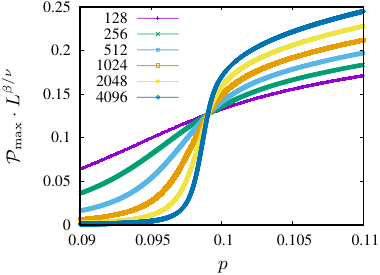}
\end{subfigure}
%% --------------------------------------------------------------
\begin{subfigure}[b]{0.4\textwidth}
\caption{\label{fig:Smax-hc-123456}} 
\includegraphics[width=0.91\textwidth]{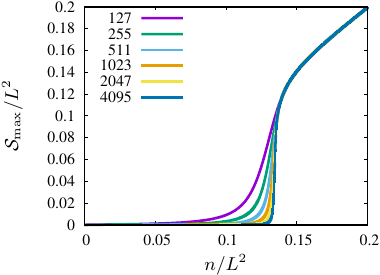}
\end{subfigure}
%% --------------------------------------------------------------
\begin{subfigure}[b]{0.4\textwidth}
\caption{\label{fig:PmaxLbetanu-hc-123456}} 
\includegraphics[width=0.91\textwidth]{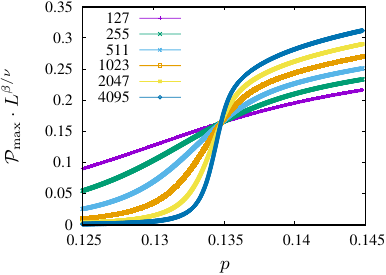}
\end{subfigure}
%% --------------------------------------------------------------
\caption{\label{fig:exmaples}Examples of [\subref{fig:Smax-tr-123456} and \subref{fig:Smax-hc-123456}] $\mathcal{S}_\text{max}/L^2$ versus $n/L^2$ and [\subref{fig:PmaxLbetanu-tr-123456} and \subref{fig:PmaxLbetanu-hc-123456}] $\mathcal{P}_{\text{max}}\cdot L^{\beta/\nu}$ versus $p$ for [\subref{fig:Smax-tr-123456} and \subref{fig:PmaxLbetanu-tr-123456}] \textsc{tr-1,2,3,4,5,6} and  [\subref{fig:Smax-hc-123456} and \subref{fig:PmaxLbetanu-hc-123456}] \textsc{hc-1,2,3,4,5,6} neighborhoods}
\end{figure}
%% ===============================================================

In \Cref{fig:exmaples} we present examples of results used to predict the percolation thresholds $p_c$, that is,
\begin{itemize}
    \item the dependencies of the size of the largest cluster $\mathcal{S}_\text{max}/L^2$ normalized to the lattice size vs. number of occupied sites also normalized to the lattice size [\Cref{fig:Smax-tr-123456,fig:Smax-hc-123456}]
    \item and the dependencies of the probability that a randomly selected site belongs to the largest cluster, scaled by $L^{\beta/\nu}$\footnote{For the problem of site percolation, the critical values of the exponents $\beta=\frac{5}{36}$ and $\nu=\frac{4}{3}$ are known exactly \cite[p.~54]{bookDS}.} vs. occupation probability $p$ [\Cref{fig:PmaxLbetanu-tr-123456,fig:PmaxLbetanu-hc-123456}]
\end{itemize}
for triangular (\Cref{fig:Smax-tr-123456,fig:PmaxLbetanu-tr-123456}) and honeycomb (\Cref{fig:Smax-hc-123456,fig:PmaxLbetanu-hc-123456}) lattice and neighbourhoods containing all considered basic neighbourhoods presented in \Cref{fig:tr-neighbors-basics} (for the triangular lattice) and \Cref{fig:hc-neighbors-basics} (for the honeycomb lattice).
The linear sizes $L$ of the simulated systems range from 127 to 4096 and the results of these simulations are averaged over $R=10^5$ samples.
All dependencies $\mathcal P_{\text{max}}\cdot L^{\beta/\nu}$ vs. $p$ studied here are presented in Figure 2 
(for the triangular lattice) and Figure 3 
(for the honeycomb lattice) in Appendix C 
in the Supplemental Material \cite{SM-EL12139}.
The common point of the curves $\mathcal{P}_\text{max}\cdot L^{\beta/\nu}$ vs. $p$ for various system sizes $L$ predicts $p_c$.
The computed values of $p_c$, associated with various neighborhoods, together with their uncertainties (also estimated earlier for neighborhoods containing sites up to the sixth coordination zone---for square lattice \cite{Galam2005b,Majewski2007,2303.10423} and the fifth coordination zone---for triangular \cite{2006.15621,2102.10066} and honeycomb \cite{2204.12593} lattices) are collected in Table I 
in Appendix B 
in the Supplemental Material \cite{SM-EL12139}. 

\Cref{fig:pc_vs_zeta} presents the $p_c$ for neighborhoods containing sites up to the sixth coordination zone on square ($\square$), honeycomb ($\fullmoon$) and triangular ($\triangle$) lattices as dependent on 
    \begin{itemize}
    \item total coordination number $z$ [\Cref{fig:pc_vs_z}];
    \item index $\zeta$ [\Cref{fig:universal}];
    \item index $\zeta/K$ [\Cref{fig:pc_vs_zetaperK}].
    \end{itemize}
The crosses ($\times$) indicates inflated neighborhoods, that is, non-compact neighborhoods reducible to other complex neighborhoods by shrinking the lattice constants. 
Three examples of inflated neighborhood are presented in \Cref{fig:equivalence}.
The detected inflated neighborhoods and their lower index equivalents are presented in \Cref{tab:inflated}.
These values $p_c$ are excluded from the fitting procedure.

%% ===============================================================
\begin{table}[htbp]
\caption{\label{tab:inflated}Detected inflated (together with the associated $\zeta$ index) and equivalent neighborhoods.
The percolation thresholds $p_c$ and total number of sites $z$ are common for both neighborhoods.}
\begin{ruledtabular}
\begin{tabular}{p{2cm} ddd p{2cm}}
%% ----------------------------------------------------------------
inflated neighborhood & \text{$\zeta$} & \text{$p_c$} & \text{$z$} & equivalent neighborhood\\ \hline
\textsc{sq-2}     &  5.6568 & 0.5927 &  4 & \textsc{sq-1}\\
\textsc{sq-3}     &  8      & 0.5927 &  4 & \textsc{sq-1}\\
\textsc{sq-5}     & 11.3137 & 0.5927 &  4 & \textsc{sq-1}\\
\textsc{sq-6}     & 12      & 0.5927 &  4 & \textsc{sq-1}\\
\textsc{sq-2,3}   & 13.6568 & 0.4073 &  8 & \textsc{sq-1,2}\\
\textsc{sq-2,5}   & 16.9705 & 0.337  &  8 & \textsc{sq-1,3}\\
\textsc{sq-3,5}   & 19.3137 & 0.4073 &  8 & \textsc{sq-1,2}\\
\textsc{sq-2,3,5} & 24.9705 & 0.288  & 12 & \textsc{sq-1,2,3}\\
\hline
\textsc{tr-2}     & 10.3923 & 0.5     &  6 & \textsc{tr-1}\\
\textsc{tr-3}     & 12      & 0.5     &  6 & \textsc{tr-1}\\
\textsc{tr-5}     & 18      & 0.5     &  6 & \textsc{tr-1}\\
\textsc{tr-6}     & 20.7846 & 0.5     &  6 & \textsc{tr-1}\\
\textsc{tr-2,5}   & 28.3923 & 0.29028 & 12 & \textsc{tr-1,2}\\
\textsc{tr-2,6}   & 31.1769 & 0.26455 & 12 & \textsc{tr-1,3}\\
\textsc{tr-3,6}   & 32.7846 & 0.29030 & 12 & \textsc{tr-1,2}\\
\textsc{tr-5,6}   & 38.7846 & 0.23200 & 12 & \textsc{tr-2,3}\\
\textsc{tr-2,5,6} & 49.1769 & 0.21550 & 18 & \textsc{tr-1,2,3}\\
\hline
\textsc{hc-2}     & 10.3923 & 0.5     &  6 & \textsc{tr-1}\\
\textsc{hc-3}     &  6      & 0.697   &  3 & \textsc{hc-1}\\
\textsc{hc-5}     & 15.5884 & 0.5     &  6 & \textsc{tr-1}\\
\textsc{hc-6}     & 20.7846 & 0.5     &  6 & \textsc{tr-1}\\
\textsc{hc-2,5}   & 28.3923 & 0.29028 & 12 & \textsc{tr-1,2}\\
\textsc{hc-2,6}   & 31.1769 & 0.26453 & 12 & \textsc{tr-1,3}\\
\textsc{hc-3,6}   & 26.7846 & 0.36301 &  9 & \textsc{hc-1,2}\\
\textsc{hc-5,6}   & 38.7846 & 0.23202 & 12 & \textsc{tr-2,3}\\
\textsc{hc-2,5,6} & 49.1769 & 0.21547 & 18 & \textsc{tr-1,2,3}\\
\end{tabular}
\end{ruledtabular}
\end{table}
%% ==============================================================

%% ==============================================================
\begin{figure*}[htbp]
%% --------------------------------------------------------------
%% --------------------------------------------------------------
\begin{subfigure}[b]{0.32\textwidth}
\caption{\label{fig:pc_vs_z}}
\includegraphics[width=0.99\columnwidth]{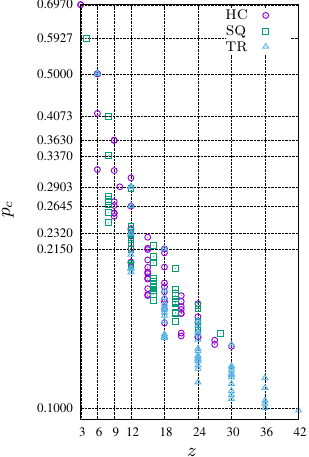}
\end{subfigure}
%% --------------------------------------------------------------
\begin{subfigure}[b]{0.32\textwidth}
\caption{\label{fig:universal}}
\includegraphics[width=0.99\columnwidth]{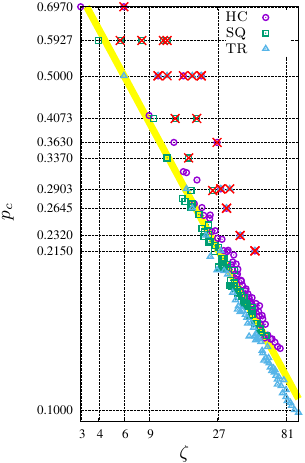}
\end{subfigure}
%% --------------------------------------------------------------
\begin{subfigure}[b]{0.32\textwidth}
\caption{\label{fig:pc_vs_zetaperK}}
\includegraphics[width=0.99\columnwidth]{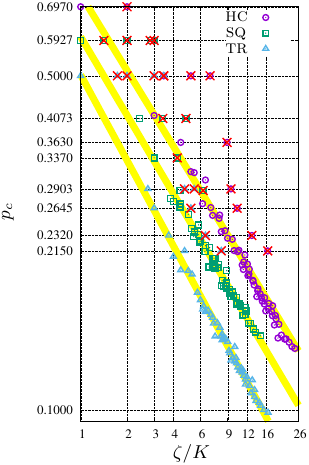}
\end{subfigure}
%% --------------------------------------------------------------
\caption{\label{fig:pc_vs_zeta}Percolation thresholds $p_c$ for neighbourhoods containing sites up to the sixth coordination zone on square ($\square$), honeycomb ($\fullmoon$) and triangular ($\triangle$) lattices as dependent on 
\subref{fig:pc_vs_z} total coordination number $z$;
\subref{fig:universal} index $\zeta$; 
\subref{fig:pc_vs_zetaperK} index $\zeta/K$.
Points marked with crosses ($\times$) correspond to inflated neighborhoods (such as those collected in \Cref{tab:inflated}), which are excluded from the fitting procedure.
The lines show power law fits according to the least-squares method to \Cref{eq:pcvszeta} and \Cref{eq:pcvszetaperK} on \Cref{fig:universal,fig:pc_vs_zetaperK}, respectively}
\end{figure*}
%% ===============================================================

%% ===============================================================
\begin{figure}[htbp]
%% ---------------------------------------------------------
\begin{subfigure}[t]{0.15\textwidth}
\caption{\label{fig:36nn-equiv-12nn}}
\includegraphics[width=\textwidth]{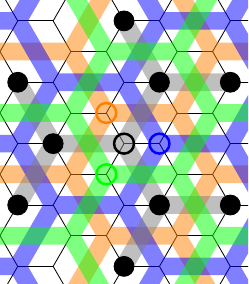}
\end{subfigure}
\hfill %% ---------------------------------------------------------
\begin{subfigure}[t]{0.15\textwidth}
\caption{\label{fig:56nn-equiv-tr23}}
\includegraphics[width=\textwidth]{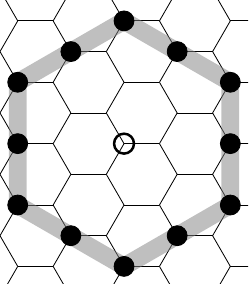}
\end{subfigure}
\hfill %% ---------------------------------------------------------
\begin{subfigure}[t]{0.15\textwidth}
\caption{\label{fig:256nn-equiv-tr123}}
\includegraphics[width=\textwidth]{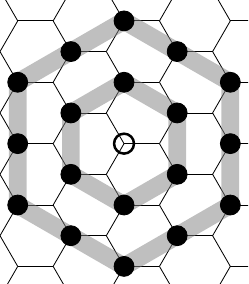}
\end{subfigure}
\hfill %% ---------------------------------------------------------
\begin{subfigure}[t]{0.15\textwidth}
\caption{\label{fig:12nn-equiv-36nn}}
\includegraphics[width=\textwidth]{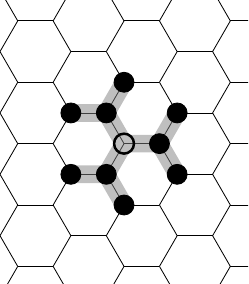}
\end{subfigure}
\hfill %% ---------------------------------------------------------
\begin{subfigure}[t]{0.15\textwidth}
\caption{\label{fig:tr23-equiv-56nn}}
\includegraphics[width=\textwidth]{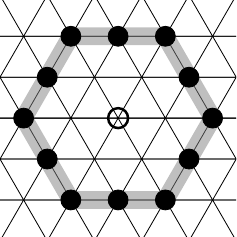}
\end{subfigure}
\hfill %% ---------------------------------------------------------
\begin{subfigure}[t]{0.15\textwidth}
\caption{\label{fig:tr123-equiv-256nn}}
\includegraphics[width=\textwidth]{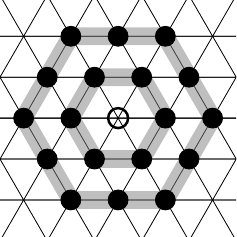}
\end{subfigure}
%% ---------------------------------------------------------
\caption{\label{fig:equivalence}Inflated complex neighbourhoods (top row) and their lower-index partners (bottom row):
\subref{fig:36nn-equiv-12nn} \textsc{hc-3,6} vs. \subref{fig:12nn-equiv-36nn} \textsc{hc-1,2},
\subref{fig:56nn-equiv-tr23} \textsc{hc-5,6} vs. \subref{fig:tr23-equiv-56nn} \textsc{tr-2,3} and
\subref{fig:256nn-equiv-tr123} \textsc{hc-2,5,6} vs. \subref{fig:tr123-equiv-256nn} \textsc{tr-1,2,3}.
The neighbourhood \textsc{hc-3,6} is equivalent to \textsc{hc-1,2} but with a lattice constant twice as large.
This split the system into four independent simultanously percolating systems}
\end{figure}
%% ===============================================================

As we mentioned in the Introduction, for complex non-compact neighborhoods, strong $p_c(z)$ degeneration is observed [see \Cref{fig:pc_vs_z}].
On the contrary, introducing the index $\zeta$ \eqref{eq:zeta} allows a nearly perfect separation of the values of $p_c$.
After excluding inflated neighborhoods (presented in \Cref{tab:inflated}) the linear fit of the data presented in \Cref{fig:pc_vs_zetaperK} with the least-squares method gives in the power law
%% --------------------------------------------------------------
\begin{equation} \label{eq:pcvszetaperK}
p_c\propto(\zeta/K)^{-\gamma} 
\end{equation}
%% --------------------------------------------------------------
exponents $\gamma_\textsc{tr}= 0.5932(47)$,
$\gamma_\textsc{sq}= 0.5439(63)$,
$\gamma_\textsc{hc}= 0.526(11)$, for triangular, square and honeycomb lattices, respectively.
The analogous fit according to \Cref{eq:pcvszeta} of the data presented in \Cref{fig:universal} gives the exponent $\gamma_\textsc{2d}=0.5546(67)$.

%% #############################################################
\section{Discussion}
%% #############################################################

The introduction of the $\zeta$ index solves the problem of multiple degeneration of the value of $p_c$.
Eliminating inflated neighborhoods (including those that occur pairwise between a triangular and a hexagonal lattice) allows fitting $p_c$ to the power laws according to Equations~\eqref{eq:pcvszeta} or~\eqref{eq:pcvszetaperK}.
Without comparing the hexagonal and triangular lattices, it was necessary to introduce the index $\xi$ to maintain the power law relationship according to Equation~\eqref{eq:pcvsxi}. The index $\xi$ turned out to be redundant vs. $\zeta$ index for the site percolation problem, because previously outlier points turned out to belong to the inflated neighborhoods, but the low-index neighborhoods associated with them were located on a different type of lattice.
However, the introduction of the index $\xi$ turned out to be quite useful for the bond percolation problem, where the relationship \eqref{eq:pcvsxi} is perfectly satisfied with the exponent $\gamma\approx 1$ \cite{Xun_2022}.

Finally, we propose some unification of the nomenclature appearing in the literature, and applying terms:
\begin{description}
\item[basic neighborhoods] for those containing sites from a single coordination zone (like \textsc{sq-1}, \textsc{sq-2}, \textsc{sq-3}, etc. and those presented in Figures~\ref{fig:tr-neighbors-basics} and \ref{fig:hc-neighbors-basics}); 
\item[complex neighborhoods] for any combination of the basic ones;
\item[extended neighborhoods] for complex and compact neighborhoods (like \textsc{sq-1,2}, \textsc{tr-1,2,3}, \textsc{hc-1,2,3,4}, etc.) and 
\item[inflated neighbourhoods] for complex neighborhoods reducible to other complex neighborhoods but with lower indexes by shrinking the lattice constant (like those presented in \Cref{fig:equivalence} and collected in Table~\ref{tab:inflated}).
\end{description}

In conclusion, in this paper we estimate percolation thresholds for the random site percolation problem on triangular and honeycomb lattices for neighborhoods containing sites from the sixth coordination zone.
The obtained values of $p_c$ satisfy the power law: independently of the underlying lattice (according to $p_c\propto\zeta^{-\gamma}$)
or even better for separately considered lattices (according to $p_c\propto(\zeta/K)^{-\gamma}$, where $K$ is the connectivity of the lattice).
Currently, the applications of complex neighborhoods on various lattice topologies seem to be most promising in agroecology \cite{ISI:000518460000003,*PhysRevE.109.014304}.

%% =============================================================
\begin{acknowledgments}
The authors thank Hubert Skawina for preparing the figures for Table I 
in Appendix B 
in the Supplemental Material \cite{SM-EL12139}.
We gratefully acknowledge Poland’s high-performance computing infrastructure PLGrid (HPC Center: ACK Cyfronet AGH) for providing computer facilities and support within computational grant no. PLG/2023/016295.
\end{acknowledgments}
%% =============================================================

%% =============================================================
%apsrev4-2.bst 2019-01-14 (MD) hand-edited version of apsrev4-1.bst
%Control: key (0)
%Control: author (8) initials jnrlst
%Control: editor formatted (1) identically to author
%Control: production of article title (0) allowed
%Control: page (0) single
%Control: year (1) truncated
%Control: production of eprint (0) enabled
%
%% =============================================================

%% ###############################################################
\appendix
%% ###############################################################

%%\begin{center}
%% \section*{Supplemental Material}
%%\end{center}

%% ===============================================================
\section{\label{app:boundaries}Procedure \texttt{boundaries()} for TR-6 and HC-6 neighborhoods}
%% ===============================================================

\Cref{lst:boundaries-tr-6,lst:boundaries-hc-6} show implementations of \texttt{boundaries()} functions to be replaced in the original Newman--Ziff algorithm.
The \texttt{boundaries()} procedures for the simple neighborhoods from \textsc{sq}-1 to \textsc{sq}-6 for square lattice are available in Appendix A of Reference~\onlinecite{2303.10423}.
The \texttt{boundaries()} procedures for the simple neighborhoods from \textsc{hc}-1 to \textsc{hc}-5 are available in Appendix A of Reference~\onlinecite{2204.12593}.

%% ==============================================================
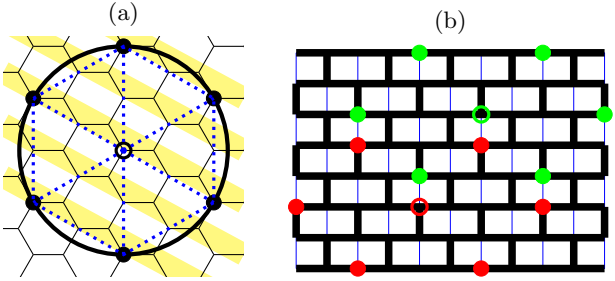
\begin{figure}[htbp]
%% --------------------------------------------------------------
%% ---------------------------------------------------------------
\begin{subfigure}[b]{0.43\columnwidth}
\caption{\label{fig:hc-6-neighbours}}
\begin{tikzpicture}[scale=0.40]
\clip (2,1) rectangle (10,9);
\draw[line width=3mm,yellow,draw opacity=.5] 
(2,13*sin{60})--(10,8*sin{60})
(2,11*sin{60})--(10,6*sin{60})
(2, 9*sin{60})--(10,4*sin{60})
(2, 7*sin{60})--(10,2*sin{60})
(2, 5*sin{60})--(10,0*sin{60});
  \foreach \i in {0,...,3} 
  \foreach \j in {0,...,5} {
  \foreach \a in {0,120,-120} \draw (3*\i,2*sin{60}*\j) -- +(\a:1);
  \foreach \a in {0,120,-120} \draw (3*\i+3*cos{60},2*sin{60}*\j+sin{60}) -- +(\a:1);}
\draw[ultra thick] (6, 6*sin{60}) circle (4*sin{60});
\draw[very thick] (6.0, 6*sin{60}) circle (7pt);
\filldraw (9, 8*sin{60}) circle (7pt);
\filldraw (9, 4*sin{60}) circle (7pt);
\filldraw (6,10*sin{60}) circle (7pt);
\filldraw (6, 2*sin{60}) circle (7pt);
\filldraw (3, 4*sin{60}) circle (7pt);
\filldraw (3, 8*sin{60}) circle (7pt);

\draw[very thick,dotted,blue]
(3, 8*sin{60})--(6,10*sin{60})--(9, 8*sin{60})--(9, 4*sin{60})--(6, 2*sin{60})--(3, 4*sin{60})--(3, 8*sin{60});

\draw[very thick,dotted,blue]
(6, 6*sin{60})--(9, 8*sin{60})
(6, 6*sin{60})--(9, 4*sin{60})
(6, 6*sin{60})--(6,10*sin{60})
(6, 6*sin{60})--(6, 2*sin{60})
(6, 6*sin{60})--(3, 4*sin{60})
(6, 6*sin{60})--(3, 8*sin{60});
\end{tikzpicture}
\end{subfigure}
%% ---------------------------------------------------------------
\begin{subfigure}[b]{0.55\columnwidth}
\caption{\label{fig:hc-6-computerized}}
\begin{tikzpicture}[scale=0.41]
\draw[step=10mm,blue,very thin] (0,2) grid (10,9);
  \foreach \j in {2,...,9} {\draw[line width=1mm, black]
  (0,\j)--(10,\j);}
  \foreach \i in {0,...,4} 
  \foreach \j in {2,4,6,8} {\draw[line width=1mm,black] (2*\i+1,\j)--(2*\i+1,\j+1);}
  \foreach \i in {0,...,5} 
  \foreach \j in {3,5,7} {\draw[line width=1mm,black]
  (2*\i,\j)--(2*\i,\j+1);}
%  \foreach \j in {2,...,9} {\draw[line width=1mm,black]
%  (0,\j)--(10,\j);}}
\draw[very thick,red]   (4,4) circle (7pt);
\filldraw[red]          (0,4) circle (7pt)
                        (8,4) circle (7pt)
                        (2,6) circle (7pt)
                        (6,6) circle (7pt)
                        (2,2) circle (7pt)
                        (6,2) circle (7pt);
\draw[very thick,green] (6,7) circle (7pt);
\filldraw[green]        (2,7) circle (7pt)
                       (10,7) circle (7pt)
                        (4,9) circle (7pt)
                        (8,9) circle (7pt)
                        (4,5) circle (7pt)
                        (8,5) circle (7pt);
\end{tikzpicture}
\end{subfigure}
%% ---------------------------------------------------------------
%% --------------------------------------------------------------
\caption{\label{fig:hc-6-neighbours-computerized}\subref{fig:hc-6-neighbours} The sixth coordination zone (\textsc{hc-6}) in the honeycomb lattice, $r^2=12$, $z=6$. 
\subref{fig:hc-6-computerized} Honeycomb lattice mapped into brick-wall like square lattice~\cite{PhysRevE.60.275}.
Sites with odd and even labels are marked as open red and green circles, respectively.
The color fulfilled circles mark sites in the neighborhood. 
The black horizontal lines correspond to yellow thick lines on panel \subref{fig:hc-6-neighbours}.
Mapping other simple neighborhoods (from \textsc{hc-1} to \textsc{hc-5}) into brick-wall like square lattice are presented in Figure 3 in Reference~\onlinecite{2204.12593}}
\end{figure}
%% =============================================================

%% --------------------------------------------------------------
\begin{lstlisting}[language=C,label=lst:boundaries-tr-6,caption=\texttt{boundaries()} for \textsc{tr-6}]
// tr-6
#define L 2048
#define Z 6
#define N (L*L)

int nn[N][Z];  /* Nearest neighbors */

void boundaries() {
int i;

for(i=0; i<N; i++) {
// tr-6 core:
  nn[i][0]=(N+i-2*L-2)%N;
  nn[i][1]=(N+i+2*L-4)%N;
  nn[i][2]=(N+i+4*L-2)%N;
  nn[i][3]=(N+i+2*L+2)%N;
  nn[i][4]=(N+i-2*L+4)%N;
  nn[i][5]=(N+i-4*L+2)%N;
// tr-6 left border:
  if(i%L==0 || i%L==1) {
    nn[i][0]=(N+L+i-2*L-2)%N;
    nn[i][1]=(N+L+i+2*L-4)%N;
    nn[i][2]=(N+L+i+4*L-2)%N; }
  if(i%L==2 || i%L==3)
    nn[i][1]=(N+L+i+2*L-4)%N;
// tr-6 right border:
  if((i+3)%L==0 || (i+4)%L==0)
    nn[i][4]=(N-L+i-2*L+4)%N;
  if((i+2)%L==0 || (i+1)%L==0) {
    nn[i][3]=(N-L+i+2*L+2)%N;
    nn[i][4]=(N-L+i-2*L+4)%N;
    nn[i][5]=(N-L+i-4*L+2)%N; }
  }
}
\end{lstlisting}
%% --------------------------------------------------------------

%% --------------------------------------------------------------
\begin{lstlisting}[language=C,label=lst:boundaries-hc-6,caption=\texttt{boundaries()} for \textsc{hc-6}]
// hc-6
#define L 2048
#define Z 6
#define N (L*L)

int nn[N][Z];  /* Nearest neighbors */

void boundaries() {
int i;

for(i=0; i<N; i++) {
// hc-6 core:
  nn[i][0] = (N+i -4)%N;
  nn[i][1] = (N+i +4)%N;
  nn[i][2] = (N+i +2*L-2)%N;
  nn[i][3] = (N+i +2*L+2)%N;
  nn[i][4] = (N+i -2*L-2)%N;
  nn[i][5] = (N+i -2*L+2)%N; 
// hc-6 left border:
  if(i%L==0 || i%L==1 || i%L==2 || i%L==3)
    nn[i][0] = (N+L+i -4)%N;
  if(i%L==0 || i%L==1) {
    nn[i][2] = (N+L+i +2*L-2)%N;
    nn[i][4] = (N+L+i -2*L-2)%N; }
// hc-6 right border:
  if((i+1)%L==0 || (i+2)%L==0 ||
     (i+3)%L==0 || (i+4)%L==0)
    nn[i][1] = (N-L+i +4)%N;
  if((i+1)%L==0 || (i+2)%L==0) {
    nn[i][3] = (N-L+i +2*L+2)%N;
    nn[i][5] = (N-L+i -2*L+2)%N; }
  }
}
\end{lstlisting}
%% --------------------------------------------------------------

%% ===============================================================
\section{\label{app:shapes}Shapes of neighborhoods and associated percolation thresholds}
%% ===============================================================

\Cref{tab:pc:hc-sq-tr} presents shapes of neighborhoods and associated percolation thresholds.

%% =============================================================
\begin{table*}[t]
\caption{\label{tab:pc:hc-sq-tr}Percolation thresholds $p_c$ for square, honeycomb and triangle lattices and neighborhoods ranging up to the sixth coordination zone}
\input{table}
\end{table*}
%% =============================================================

%% ===============================================================
\section{\label{app:PmaxLbetanuvsp}Scaled probability of belonging to the largest cluster vs. occupation probability}
%% ===============================================================

\Cref{fig:PmaxLbetanuvsp-tr,fig:PmaxLbetanuvsp-hc} shows scaled (by factor $L^{\beta/\nu}$) probability $\mathcal P_{\text{max}}$ of belonging to the largest cluster vs. the probability of occupation $p$.

%% ===============================================================
\begin{figure*}[htbp]
%% ---------------------------------------------------------------
\begin{subfigure}[b]{0.27\textwidth}
%% \psfrag{PmaxLx}[][c]{$\mathcal{P}_{\max}\cdot L^{\beta/\nu}$}
%% \psfrag{p}{$p$}
%% \psfrag{L}{$L=$}
\caption{\label{fig:PmaxLbetanuvsp-tr-6}\textsc{tr-6}}
\includegraphics[width=.99\textwidth]{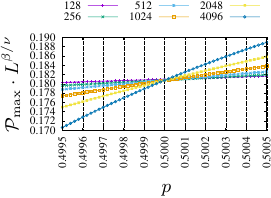}
\end{subfigure}
\hfill %% --------------------------------------------------------
\begin{subfigure}[b]{0.27\textwidth}
%% \psfrag{PmaxLx}[][c]{$\mathcal{P}_{\max}\cdot L^{\beta/\nu}$}
%% \psfrag{p}{$p$}
%% \psfrag{L}{$L=$}
\caption{\label{fig:PmaxLbetanuvsp-tr-16}\textsc{tr-1,6}}
\includegraphics[width=.99\textwidth]{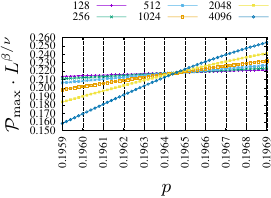}
\end{subfigure}
\hfill %% --------------------------------------------------------
\begin{subfigure}[b]{0.27\textwidth}
%% \psfrag{PmaxLx}[][c]{$\mathcal{P}_{\max}\cdot L^{\beta/\nu}$}
%% \psfrag{p}{$p$}
%% \psfrag{L}{$L=$}
\caption{\label{fig:PmaxLbetanuvsp-tr-26}\textsc{tr-2,6}}
\includegraphics[width=.99\textwidth]{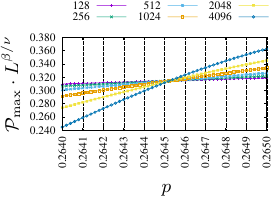}
\end{subfigure}
\hfill %% --------------------------------------------------------
\begin{subfigure}[b]{0.27\textwidth}
%% \psfrag{PmaxLx}[][c]{$\mathcal{P}_{\max}\cdot L^{\beta/\nu}$}
%% \psfrag{p}{$p$}
%% \psfrag{L}{$L=$}
\caption{\label{fig:PmaxLbetanuvsp-tr-36}\textsc{tr-3,6}}
\includegraphics[width=.99\textwidth]{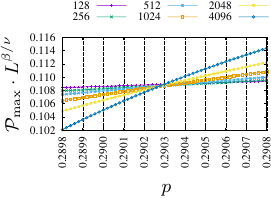}
\end{subfigure}
\hfill %% --------------------------------------------------------
\begin{subfigure}[b]{0.27\textwidth}
%% \psfrag{PmaxLx}[][c]{$\mathcal{P}_{\max}\cdot L^{\beta/\nu}$}
%% \psfrag{p}{$p$}
%% \psfrag{L}{$L=$}
\caption{\label{fig:PmaxLbetanuvsp-tr-46}\textsc{tr-4,6}}
\includegraphics[width=.99\textwidth]{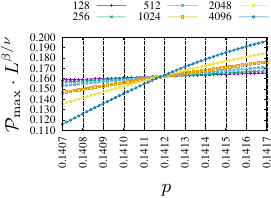}
\end{subfigure}
\hfill %% --------------------------------------------------------
\begin{subfigure}[b]{0.27\textwidth}
%% \psfrag{PmaxLx}[][c]{$\mathcal{P}_{\max}\cdot L^{\beta/\nu}$}
%% \psfrag{p}{$p$}
%% \psfrag{L}{$L=$}
\caption{\label{fig:PmaxLbetanuvsp-tr-56}\textsc{tr-5,6}}
\includegraphics[width=.99\textwidth]{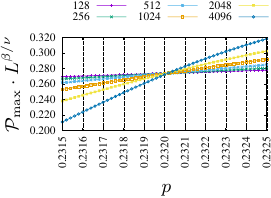}
\end{subfigure}
\hfill %% --------------------------------------------------------
\begin{subfigure}[b]{0.27\textwidth}
%% \psfrag{PmaxLx}[][c]{$\mathcal{P}_{\max}\cdot L^{\beta/\nu}$}
%% \psfrag{p}{$p$}
%% \psfrag{L}{$L=$}
\caption{\label{fig:PmaxLbetanuvsp-tr-126}\textsc{tr-1,2,6}}
\includegraphics[width=.99\textwidth]{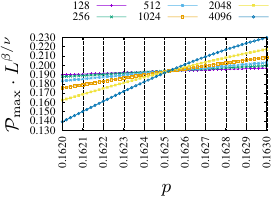}
\end{subfigure}
\hfill %% --------------------------------------------------------
\begin{subfigure}[b]{0.27\textwidth}
%% \psfrag{PmaxLx}[][c]{$\mathcal{P}_{\max}\cdot L^{\beta/\nu}$}
%% \psfrag{p}{$p$}
%% \psfrag{L}{$L=$}
\caption{\label{fig:PmaxLbetanuvsp-tr-136}\textsc{tr-1,3,6}}
\includegraphics[width=.99\textwidth]{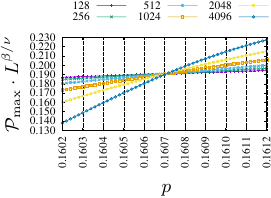}
\end{subfigure}
\hfill %% --------------------------------------------------------
\begin{subfigure}[b]{0.27\textwidth}
%% \psfrag{PmaxLx}[][c]{$\mathcal{P}_{\max}\cdot L^{\beta/\nu}$}
%% \psfrag{p}{$p$}
%% \psfrag{L}{$L=$}
\caption{\label{fig:PmaxLbetanuvsp-tr-146}\textsc{tr-1,4,6}}
\includegraphics[width=.99\textwidth]{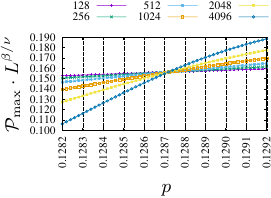}
\end{subfigure}
\hfill %% --------------------------------------------------------
\begin{subfigure}[b]{0.27\textwidth}
%% \psfrag{PmaxLx}[][c]{$\mathcal{P}_{\max}\cdot L^{\beta/\nu}$}
%% \psfrag{p}{$p$}
%% \psfrag{L}{$L=$}
\caption{\label{fig:PmaxLbetanuvsp-tr-156}\textsc{tr-1,5,6}}
\includegraphics[width=.99\textwidth]{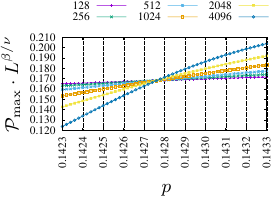}
\end{subfigure}
\hfill %% --------------------------------------------------------
\begin{subfigure}[b]{0.27\textwidth}
%% \psfrag{PmaxLx}[][c]{$\mathcal{P}_{\max}\cdot L^{\beta/\nu}$}
%% \psfrag{p}{$p$}
%% \psfrag{L}{$L=$}
\caption{\label{fig:PmaxLbetanuvsp-tr-236}\textsc{tr-2,3,6}}
\includegraphics[width=.99\textwidth]{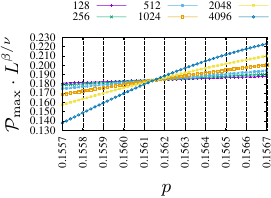}
\end{subfigure}
\hfill %% --------------------------------------------------------
\begin{subfigure}[b]{0.27\textwidth}
%% \psfrag{PmaxLx}[][c]{$\mathcal{P}_{\max}\cdot L^{\beta/\nu}$}
%% \psfrag{p}{$p$}
%% \psfrag{L}{$L=$}
\caption{\label{fig:PmaxLbetanuvsp-tr-246}\textsc{tr-2,4,6}}
\includegraphics[width=.99\textwidth]{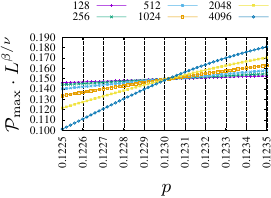}
\end{subfigure}
\hfill %% --------------------------------------------------------
\begin{subfigure}[b]{0.27\textwidth}
%% \psfrag{PmaxLx}[][c]{$\mathcal{P}_{\max}\cdot L^{\beta/\nu}$}
%% \psfrag{p}{$p$}
%% \psfrag{L}{$L=$}
\caption{\label{fig:PmaxLbetanuvsp-tr-256}\textsc{tr-2,5,6}}
\includegraphics[width=.99\textwidth]{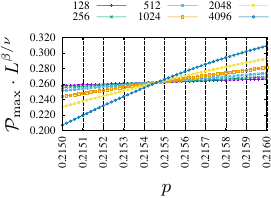}
\end{subfigure}
\hfill %% --------------------------------------------------------
\begin{subfigure}[b]{0.27\textwidth}
%% \psfrag{PmaxLx}[][c]{$\mathcal{P}_{\max}\cdot L^{\beta/\nu}$}
%% \psfrag{p}{$p$}
%% \psfrag{L}{$L=$}
\caption{\label{fig:PmaxLbetanuvsp-tr-346}\textsc{tr-3,4,6}}
\includegraphics[width=.99\textwidth]{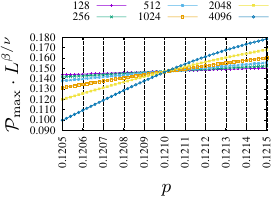}
\end{subfigure}
\hfill %% --------------------------------------------------------
\begin{subfigure}[b]{0.27\textwidth}
%% \psfrag{PmaxLx}[][c]{$\mathcal{P}_{\max}\cdot L^{\beta/\nu}$}
%% \psfrag{p}{$p$}
%% \psfrag{L}{$L=$}
\caption{\label{fig:PmaxLbetanuvsp-tr-356}\textsc{tr-3,5,6}}
\includegraphics[width=.99\textwidth]{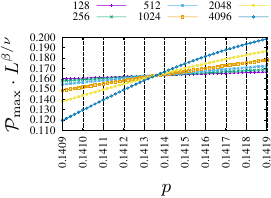}
\end{subfigure}
\hfill %% -------------------------------------------------------
\begin{subfigure}[b]{0.27\textwidth}
%% \psfrag{PmaxLx}[][c]{$\mathcal{P}_{\max}\cdot L^{\beta/\nu}$}
%% \psfrag{p}{$p$}
%% \psfrag{L}{$L=$}
\caption{\label{fig:PmaxLbetanuvsp-tr-456}\textsc{tr-4,5,6}}
\includegraphics[width=.99\textwidth]{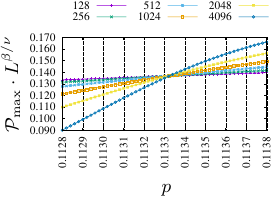}
\end{subfigure}
\hfill %% --------------------------------------------------------
\begin{subfigure}[b]{0.27\textwidth}
%% \psfrag{PmaxLx}[][c]{$\mathcal{P}_{\max}\cdot L^{\beta/\nu}$}
%% \psfrag{p}{$p$}
%% \psfrag{L}{$L=$}
\caption{\label{fig:PmaxLbetanuvsp-tr-1236}\textsc{tr-1,2,3,6}}
\includegraphics[width=.99\textwidth]{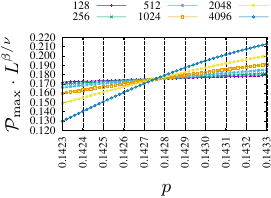}
\end{subfigure}
\hfill %% --------------------------------------------------------
\begin{subfigure}[b]{0.27\textwidth}
%% \psfrag{PmaxLx}[][c]{$\mathcal{P}_{\max}\cdot L^{\beta/\nu}$}
%% \psfrag{p}{$p$}
%% \psfrag{L}{$L=$}
\caption{\label{fig:PmaxLbetanuvsp-tr-1246}\textsc{tr-1,2,4,6}}
\includegraphics[width=.99\textwidth]{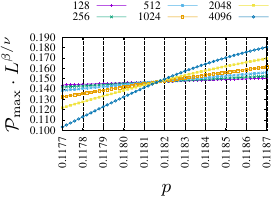}
\end{subfigure}
%% ---------------------------------------------------------------
\end{figure*}
%% +++++++++++++++++++++++++++++++++++++++++++++++++++++++++++++++
\begin{figure*}\ContinuedFloat
%% ---------------------------------------------------------------
\begin{subfigure}[b]{0.27\textwidth}
%% \psfrag{PmaxLx}[][c]{$\mathcal{P}_{\max}\cdot L^{\beta/\nu}$}
%% \psfrag{p}{$p$}
%% \psfrag{L}{$L=$}
\caption{\label{fig:PmaxLbetanuvsp-tr-1256}\textsc{tr-1,2,5,6}}
\includegraphics[width=.99\textwidth]{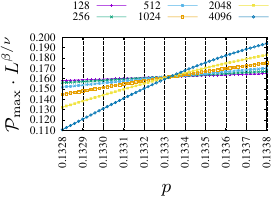}
\end{subfigure}
\hfill %% --------------------------------------------------------
\begin{subfigure}[b]{0.27\textwidth}
%% \psfrag{PmaxLx}[][c]{$\mathcal{P}_{\max}\cdot L^{\beta/\nu}$}
%% \psfrag{p}{$p$}
%% \psfrag{L}{$L=$}
\caption{\label{fig:PmaxLbetanuvsp-tr-1346}\textsc{tr-1,3,4,6}}
\includegraphics[width=.99\textwidth]{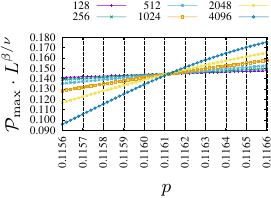}
\end{subfigure}
\hfill %% --------------------------------------------------------
\begin{subfigure}[b]{0.27\textwidth}
%% \psfrag{PmaxLx}[][c]{$\mathcal{P}_{\max}\cdot L^{\beta/\nu}$}
%% \psfrag{p}{$p$}
%% \psfrag{L}{$L=$}
\caption{\label{fig:PmaxLbetanuvsp-tr-1356}\textsc{tr-1,3,5,6}}
\includegraphics[width=.99\textwidth]{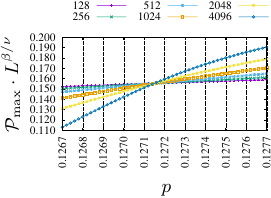}
\end{subfigure}
\hfill %% --------------------------------------------------------
\begin{subfigure}[b]{0.27\textwidth}
%% \psfrag{PmaxLx}[][c]{$\mathcal{P}_{\max}\cdot L^{\beta/\nu}$}
%% \psfrag{p}{$p$}
%% \psfrag{L}{$L=$}
\caption{\label{fig:PmaxLbetanuvsp-tr-1456}\textsc{tr-1,4,5,6}}
\includegraphics[width=.99\textwidth]{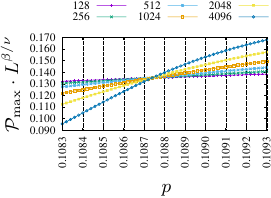}
\end{subfigure}
\hfill %% --------------------------------------------------------
\begin{subfigure}[b]{0.27\textwidth}
%% \psfrag{PmaxLx}[][c]{$\mathcal{P}_{\max}\cdot L^{\beta/\nu}$}
%% \psfrag{p}{$p$}
%% \psfrag{L}{$L=$}
\caption{\label{fig:PmaxLbetanuvsp-tr-2346}\textsc{tr-2,3,4,6}}
\includegraphics[width=.99\textwidth]{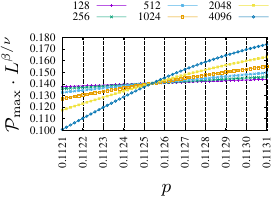}
\end{subfigure}
\hfill %% --------------------------------------------------------
\begin{subfigure}[b]{0.27\textwidth}
%% \psfrag{PmaxLx}[][c]{$\mathcal{P}_{\max}\cdot L^{\beta/\nu}$}
%% \psfrag{p}{$p$}
%% \psfrag{L}{$L=$}
\caption{\label{fig:PmaxLbetanuvsp-tr-2356}\textsc{tr-2,3,5,6}}
\includegraphics[width=.99\textwidth]{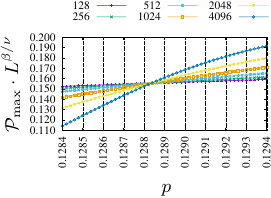}
\end{subfigure}
\hfill %% --------------------------------------------------------
\begin{subfigure}[b]{0.27\textwidth}
%% \psfrag{PmaxLx}[][c]{$\mathcal{P}_{\max}\cdot L^{\beta/\nu}$}
%% \psfrag{p}{$p$}
%% \psfrag{L}{$L=$}
\caption{\label{fig:PmaxLbetanuvsp-tr-2456}\textsc{tr-2,4,5,6}}
\includegraphics[width=.99\textwidth]{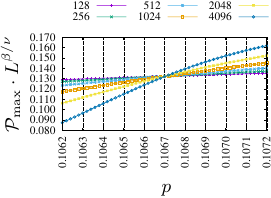}
\end{subfigure}
\hfill %% --------------------------------------------------------
\begin{subfigure}[b]{0.27\textwidth}
%% \psfrag{PmaxLx}[][c]{$\mathcal{P}_{\max}\cdot L^{\beta/\nu}$}
%% \psfrag{p}{$p$}
%% \psfrag{L}{$L=$}
\caption{\label{fig:PmaxLbetanuvsp-tr-3456}\textsc{tr-3,4,5,6}}
\includegraphics[width=.99\textwidth]{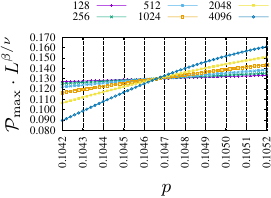}
\end{subfigure}
\hfill %% --------------------------------------------------------
\begin{subfigure}[b]{0.27\textwidth}
%% \psfrag{PmaxLx}[][c]{$\mathcal{P}_{\max}\cdot L^{\beta/\nu}$}
%% \psfrag{p}{$p$}
%% \psfrag{L}{$L=$}
\caption{\label{fig:PmaxLbetanuvsp-tr-12346}\textsc{tr-1,2,3,4,6}}
\includegraphics[width=.99\textwidth]{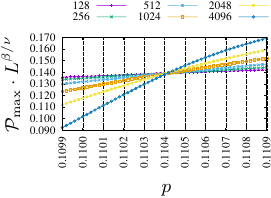}
\end{subfigure}
\hfill %% --------------------------------------------------------
\begin{subfigure}[b]{0.27\textwidth}
%% \psfrag{PmaxLx}[][c]{$\mathcal{P}_{\max}\cdot L^{\beta/\nu}$}
%% \psfrag{p}{$p$}
%% \psfrag{L}{$L=$}
\caption{\label{fig:PmaxLbetanuvsp-tr-12356}\textsc{tr-1,2,3,5,6}}
\includegraphics[width=.99\textwidth]{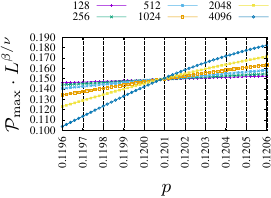}
\end{subfigure}
\hfill %% --------------------------------------------------------
\begin{subfigure}[b]{0.27\textwidth}
%% \psfrag{PmaxLx}[][c]{$\mathcal{P}_{\max}\cdot L^{\beta/\nu}$}
%% \psfrag{p}{$p$}
%% \psfrag{L}{$L=$}
\caption{\label{fig:PmaxLbetanuvsp-tr-12456}\textsc{tr-1,2,4,5,6}}
\includegraphics[width=.99\textwidth]{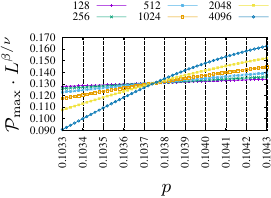}
\end{subfigure}
\hfill %% --------------------------------------------------------
\begin{subfigure}[b]{0.27\textwidth}
%% \psfrag{PmaxLx}[][c]{$\mathcal{P}_{\max}\cdot L^{\beta/\nu}$}
%% \psfrag{p}{$p$}
%% \psfrag{L}{$L=$}
\caption{\label{fig:PmaxLbetanuvsp-tr-13456}\textsc{tr-1,3,4,5,6}}
\includegraphics[width=.99\textwidth]{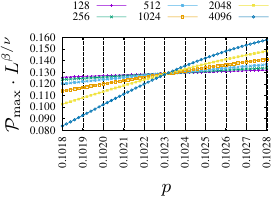}
\end{subfigure}
\hfill %% --------------------------------------------------------
\begin{subfigure}[b]{0.27\textwidth}
%% \psfrag{PmaxLx}[][c]{$\mathcal{P}_{\max}\cdot L^{\beta/\nu}$}
%% \psfrag{p}{$p$}
%% \psfrag{L}{$L=$}
\caption{\label{fig:PmaxLbetanuvsp-tr-23456}\textsc{tr-2,3,4,5,6}}
\includegraphics[width=.99\textwidth]{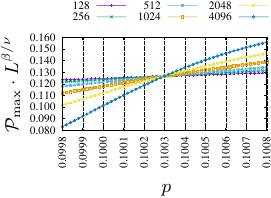}
\end{subfigure}
\hfill %% --------------------------------------------------------
\begin{subfigure}[b]{0.27\textwidth}
%% \psfrag{PmaxLx}[][c]{$\mathcal{P}_{\max}\cdot L^{\beta/\nu}$}
%% \psfrag{p}{$p$}
%% \psfrag{L}{$L=$}
\caption{\label{fig:PmaxLbetanuvsp-tr-123456}\textsc{tr-1,2,3,4,5,6}}
\includegraphics[width=.99\textwidth]{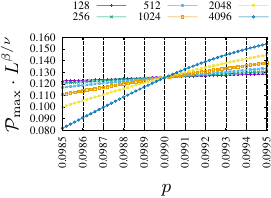}
\end{subfigure}
%% ---------------------------------------------------------------
\caption{\label{fig:PmaxLbetanuvsp-tr}$\mathcal{P}_{\max}\cdot L^{\beta/\nu}$ vs. $p$ for various complex neighborhoods on triangular lattice. 
Results are averaged over $R=10^5$ simulations and $\Delta p=2\times 10^{-5}$.
\subref{fig:PmaxLbetanuvsp-tr-6}      \textsc{tr-6},
\subref{fig:PmaxLbetanuvsp-tr-16}     \textsc{tr-1,6},
\subref{fig:PmaxLbetanuvsp-tr-26}     \textsc{tr-2,6},
\subref{fig:PmaxLbetanuvsp-tr-36}     \textsc{tr-3,6},
\subref{fig:PmaxLbetanuvsp-tr-46}     \textsc{tr-4,6},
\subref{fig:PmaxLbetanuvsp-tr-56}     \textsc{tr-5,6},
\subref{fig:PmaxLbetanuvsp-tr-126}    \textsc{tr-1,2,6},
\subref{fig:PmaxLbetanuvsp-tr-136}    \textsc{tr-1,3,6},
\subref{fig:PmaxLbetanuvsp-tr-146}    \textsc{tr-1,4,6},
\subref{fig:PmaxLbetanuvsp-tr-156}    \textsc{tr-1,5,6},
\subref{fig:PmaxLbetanuvsp-tr-236}    \textsc{tr-2,3,6},
\subref{fig:PmaxLbetanuvsp-tr-246}    \textsc{tr-2,4,6},
\subref{fig:PmaxLbetanuvsp-tr-256}    \textsc{tr-2,5,6},
\subref{fig:PmaxLbetanuvsp-tr-346}    \textsc{tr-3,4,6},
\subref{fig:PmaxLbetanuvsp-tr-356}    \textsc{tr-3,5,6},
\subref{fig:PmaxLbetanuvsp-tr-456}    \textsc{tr-4,5,6},
\subref{fig:PmaxLbetanuvsp-tr-1236}   \textsc{tr-1,2,3,6},
\subref{fig:PmaxLbetanuvsp-tr-1246}   \textsc{tr-1,2,4,6},
\subref{fig:PmaxLbetanuvsp-tr-1256}   \textsc{tr-1,2,5,6},
\subref{fig:PmaxLbetanuvsp-tr-1346}   \textsc{tr-1,3,4,6},
\subref{fig:PmaxLbetanuvsp-tr-1356}   \textsc{tr-1,3,5,6},
\subref{fig:PmaxLbetanuvsp-tr-1456}   \textsc{tr-1,4,5,6},
\subref{fig:PmaxLbetanuvsp-tr-2346}   \textsc{tr-2,3,4,6},
\subref{fig:PmaxLbetanuvsp-tr-2356}   \textsc{tr-2,3,5,6},
\subref{fig:PmaxLbetanuvsp-tr-2456}   \textsc{tr-2,4,5,6},
\subref{fig:PmaxLbetanuvsp-tr-3456}   \textsc{tr-3,4,5,6},
\subref{fig:PmaxLbetanuvsp-tr-12346}  \textsc{tr-1,2,3,4,6},
\subref{fig:PmaxLbetanuvsp-tr-12356}  \textsc{tr-1,2,3,5,6},
\subref{fig:PmaxLbetanuvsp-tr-12456}  \textsc{tr-1,2,4,5,6},
\subref{fig:PmaxLbetanuvsp-tr-13456}  \textsc{tr-1,3,4,5,6},
\subref{fig:PmaxLbetanuvsp-tr-23456}  \textsc{tr-2,3,4,5,6},
\subref{fig:PmaxLbetanuvsp-tr-123456} \textsc{tr-1,2,3,4,5,6}}
\end{figure*}
%% ===============================================================

%% ===============================================================
\begin{figure*}[htbp]
%% ---------------------------------------------------------------
\begin{subfigure}[b]{0.27\textwidth}
%% \psfrag{PmaxLx}[][c]{$\mathcal{P}_{\max}\cdot L^{\beta/\nu}$}
%% \psfrag{p}{$p$}
%% \psfrag{L}{$L=$}
\caption{\label{fig:PmaxLbetanuvsp-hc-6}\textsc{hc}-6}
\includegraphics[width=.99\textwidth]{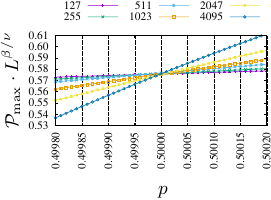}
\end{subfigure}
\hfill %% --------------------------------------------------------
\begin{subfigure}[b]{0.27\textwidth}
%% \psfrag{PmaxLx}[][c]{$\mathcal{P}_{\max}\cdot L^{\beta/\nu}$}
%% \psfrag{p}{$p$}
%% \psfrag{L}{$L=$}
\caption{\label{fig:PmaxLbetanuvsp-hc-16}\textsc{hc}-1,6}
\includegraphics[width=.99\textwidth]{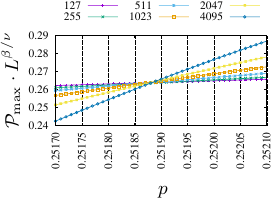}
\end{subfigure}
\hfill %% --------------------------------------------------------
\begin{subfigure}[b]{0.27\textwidth}
%% \psfrag{PmaxLx}[][c]{$\mathcal{P}_{\max}\cdot L^{\beta/\nu}$}
%% \psfrag{p}{$p$}
%% \psfrag{L}{$L=$}
\caption{\label{fig:PmaxLbetanuvsp-hc-26}\textsc{hc}-2,6}
\includegraphics[width=.99\textwidth]{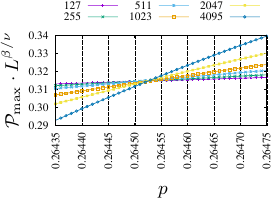}
\end{subfigure}
\hfill %% --------------------------------------------------------
\begin{subfigure}[b]{0.27\textwidth}
%% \psfrag{PmaxLx}[][c]{$\mathcal{P}_{\max}\cdot L^{\beta/\nu}$}
%% \psfrag{p}{$p$}
%% \psfrag{L}{$L=$}
\caption{\label{fig:PmaxLbetanuvsp-hc-36}\textsc{hc-3,6}} %%  $\equiv$ \textsc{hc-1,2}
\includegraphics[width=.99\textwidth]{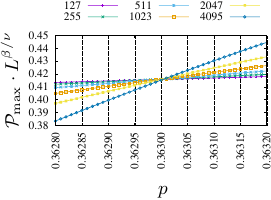}
\end{subfigure}
\hfill %% --------------------------------------------------------
\begin{subfigure}[b]{0.27\textwidth}
%% \psfrag{PmaxLx}[][c]{$\mathcal{P}_{\max}\cdot L^{\beta/\nu}$}
%% \psfrag{p}{$p$}
%% \psfrag{L}{$L=$}
\caption{\label{fig:PmaxLbetanuvsp-hc-46}\textsc{hc}-4,6}
\includegraphics[width=.99\textwidth]{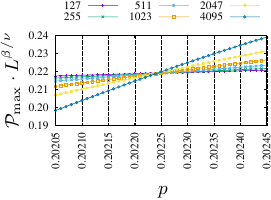}
\end{subfigure}
\hfill %% --------------------------------------------------------
\begin{subfigure}[b]{0.27\textwidth}
%% \psfrag{PmaxLx}[][c]{$\mathcal{P}_{\max}\cdot L^{\beta/\nu}$}
%% \psfrag{p}{$p$}
%% \psfrag{L}{$L=$}
\caption{\label{fig:PmaxLbetanuvsp-hc-56}\textsc{hc}-5,6}
\includegraphics[width=.99\textwidth]{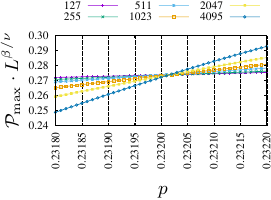}
\end{subfigure}
\hfill %% --------------------------------------------------------
\begin{subfigure}[b]{0.27\textwidth}
%% \psfrag{PmaxLx}[][c]{$\mathcal{P}_{\max}\cdot L^{\beta/\nu}$}
%% \psfrag{p}{$p$}
%% \psfrag{L}{$L=$}
\caption{\label{fig:PmaxLbetanuvsp-hc-126}\textsc{hc}-1,2,6}
\includegraphics[width=.99\textwidth]{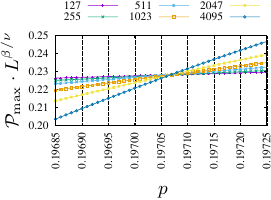}
\end{subfigure}
\hfill %% --------------------------------------------------------
\begin{subfigure}[b]{0.27\textwidth}
%% \psfrag{PmaxLx}[][c]{$\mathcal{P}_{\max}\cdot L^{\beta/\nu}$}
%% \psfrag{p}{$p$}
%% \psfrag{L}{$L=$}
\caption{\label{fig:PmaxLbetanuvsp-hc-136}\textsc{hc}-1,3,6}
\includegraphics[width=.99\textwidth]{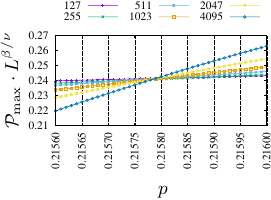}
\end{subfigure}
\hfill %% --------------------------------------------------------
\begin{subfigure}[b]{0.27\textwidth}
%% \psfrag{PmaxLx}[][c]{$\mathcal{P}_{\max}\cdot L^{\beta/\nu}$}
%% \psfrag{p}{$p$}
%% \psfrag{L}{$L=$}
\caption{\label{fig:PmaxLbetanuvsp-hc-146}\textsc{hc}-1,4,6}
\includegraphics[width=.99\textwidth]{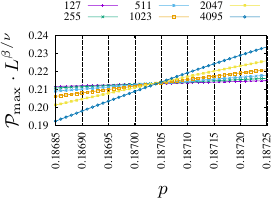}
\end{subfigure}
\hfill %% --------------------------------------------------------
\begin{subfigure}[b]{0.27\textwidth}
%% \psfrag{PmaxLx}[][c]{$\mathcal{P}_{\max}\cdot L^{\beta/\nu}$}
%% \psfrag{p}{$p$}
%% \psfrag{L}{$L=$}
\caption{\label{fig:PmaxLbetanuvsp-hc-156}\textsc{hc}-1,5,6}
\includegraphics[width=.99\textwidth]{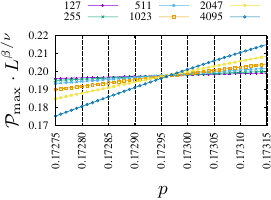}
\end{subfigure}
\hfill %% --------------------------------------------------------
\begin{subfigure}[b]{0.27\textwidth}
%% \psfrag{PmaxLx}[][c]{$\mathcal{P}_{\max}\cdot L^{\beta/\nu}$}
%% \psfrag{p}{$p$}
%% \psfrag{L}{$L=$}
\caption{\label{fig:PmaxLbetanuvsp-hc-236}\textsc{hc}-2,3,6}
\includegraphics[width=.99\textwidth]{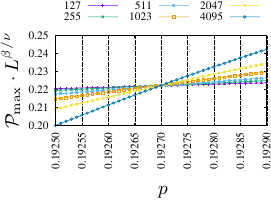}
\end{subfigure}
\hfill %% --------------------------------------------------------
\begin{subfigure}[b]{0.27\textwidth}
%% \psfrag{PmaxLx}[][c]{$\mathcal{P}_{\max}\cdot L^{\beta/\nu}$}
%% \psfrag{p}{$p$}
%% \psfrag{L}{$L=$}
\caption{\label{fig:PmaxLbetanuvsp-hc-246}\textsc{hc}-2,4,6}
\includegraphics[width=.99\textwidth]{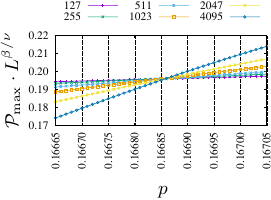}
\end{subfigure}
\hfill %% --------------------------------------------------------
\begin{subfigure}[b]{0.27\textwidth}
%% \psfrag{PmaxLx}[][c]{$\mathcal{P}_{\max}\cdot L^{\beta/\nu}$}
%% \psfrag{p}{$p$}
%% \psfrag{L}{$L=$}
\caption{\label{fig:PmaxLbetanuvsp-hc-256}\textsc{hc}-2,5,6}
\includegraphics[width=.99\textwidth]{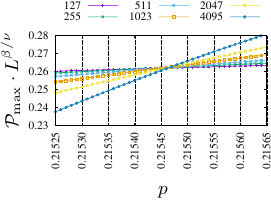}
\end{subfigure}
\hfill %% --------------------------------------------------------
\begin{subfigure}[b]{0.27\textwidth}
%% \psfrag{PmaxLx}[][c]{$\mathcal{P}_{\max}\cdot L^{\beta/\nu}$}
%% \psfrag{p}{$p$}
%% \psfrag{L}{$L=$}
\caption{\label{fig:PmaxLbetanuvsp-hc-346}\textsc{hc}-3,4,6}
\includegraphics[width=.99\textwidth]{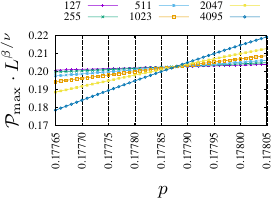}
\end{subfigure}
\hfill %% --------------------------------------------------------
\begin{subfigure}[b]{0.27\textwidth}
%% \psfrag{PmaxLx}[][c]{$\mathcal{P}_{\max}\cdot L^{\beta/\nu}$}
%% \psfrag{p}{$p$}
%% \psfrag{L}{$L=$}
\caption{\label{fig:PmaxLbetanuvsp-hc-356}\textsc{hc}-3,5,6}
\includegraphics[width=.99\textwidth]{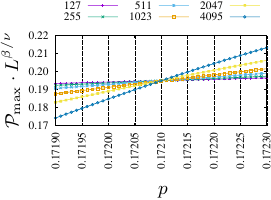}
\end{subfigure}
\hfill %% --------------------------------------------------------
\begin{subfigure}[b]{0.27\textwidth}
%% \psfrag{PmaxLx}[][c]{$\mathcal{P}_{\max}\cdot L^{\beta/\nu}$}
%% \psfrag{p}{$p$}
%% \psfrag{L}{$L=$}
\caption{\label{fig:PmaxLbetanuvsp-hc-456}\textsc{hc-4,5,6}}
\includegraphics[width=.99\textwidth]{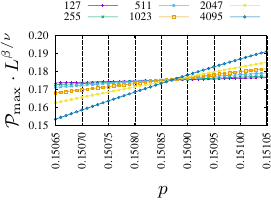}
\end{subfigure}
\hfill %% --------------------------------------------------------
\begin{subfigure}[b]{0.27\textwidth}
%%\psfrag{PmaxLx}[][c]{$\mathcal{P}_{\max}\cdot L^{\beta/\nu}$}
%%\psfrag{p}{$p$}
%%\psfrag{L}{$L=$}
\caption{\label{fig:PmaxLbetanuvsp-hc-1236}\textsc{hc-1,2,3,6}}
\includegraphics[width=.99\textwidth]{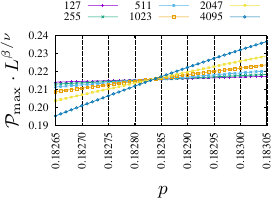}
\end{subfigure}
\hfill %% --------------------------------------------------------
\begin{subfigure}[b]{0.27\textwidth}
%% \psfrag{PmaxLx}[][c]{$\mathcal{P}_{\max}\cdot L^{\beta/\nu}$}
%% \psfrag{p}{$p$}
%% \psfrag{L}{$L=$}
\caption{\label{fig:PmaxLbetanuvsp-hc-1246}\textsc{hc-1,2,4,6}}
\includegraphics[width=.99\textwidth]{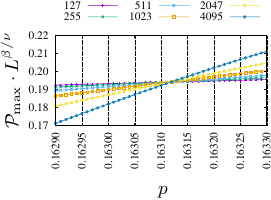}
\end{subfigure}
%% ---------------------------------------------------------------
\end{figure*}
%% +++++++++++++++++++++++++++++++++++++++++++++++++++++++++++++++
\begin{figure*}\ContinuedFloat
%% ---------------------------------------------------------------
\begin{subfigure}[b]{0.27\textwidth}
%%\psfrag{PmaxLx}[][c]{$\mathcal{P}_{\max}\cdot L^{\beta/\nu}$}
%%\psfrag{p}{$p$}
%%\psfrag{L}{$L=$}
\caption{\label{fig:PmaxLbetanuvsp-hc-1256}\textsc{hc-1,2,5,6}}
\includegraphics[width=.99\textwidth]{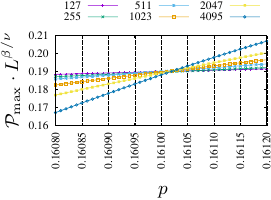}
\end{subfigure}
\hfill %% --------------------------------------------------------
\begin{subfigure}[b]{0.27\textwidth}
%%\psfrag{PmaxLx}[][c]{$\mathcal{P}_{\max}\cdot L^{\beta/\nu}$}
%%\psfrag{p}{$p$}
%%\psfrag{L}{$L=$}
\caption{\label{fig:PmaxLbetanuvsp-hc-1346}\textsc{hc-1,3,4,6}}
\includegraphics[width=.99\textwidth]{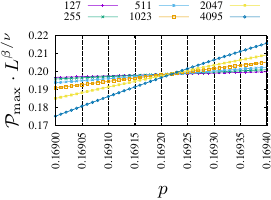}
\end{subfigure}
\hfill %% --------------------------------------------------------
\begin{subfigure}[b]{0.27\textwidth}
%%\psfrag{PmaxLx}[][c]{$\mathcal{P}_{\max}\cdot L^{\beta/\nu}$}
%%\psfrag{p}{$p$}
%%\psfrag{L}{$L=$}
\caption{\label{fig:PmaxLbetanuvsp-hc-1356}\textsc{hc-1,3,5,6}}
\includegraphics[width=.99\textwidth]{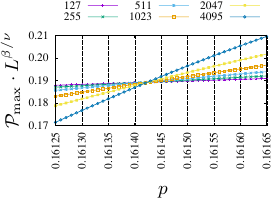}
\end{subfigure}
\hfill %% --------------------------------------------------------
\begin{subfigure}[b]{0.27\textwidth}
%% \psfrag{PmaxLx}[][c]{$\mathcal{P}_{\max}\cdot L^{\beta/\nu}$}
%% \psfrag{p}{$p$}
%% \psfrag{L}{$L=$}
\caption{\label{fig:PmaxLbetanuvsp-hc-1456}\textsc{hc-1,4,5,6}}
\includegraphics[width=.99\textwidth]{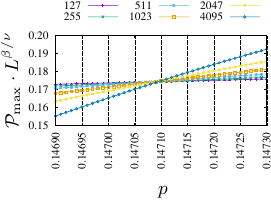}
\end{subfigure}
\hfill %% --------------------------------------------------------
\begin{subfigure}[b]{0.27\textwidth}
%% \psfrag{PmaxLx}[][c]{$\mathcal{P}_{\max}\cdot L^{\beta/\nu}$}
%% \psfrag{p}{$p$}
%% \psfrag{L}{$L=$}
\caption{\label{fig:PmaxLbetanuvsp-hc-2346}\textsc{hc-2,3,4,6}}
\includegraphics[width=.99\textwidth]{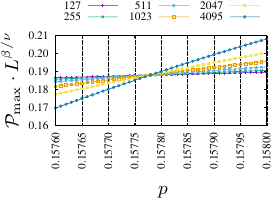}
\end{subfigure}
\hfill %% --------------------------------------------------------
\begin{subfigure}[b]{0.27\textwidth}
%% \psfrag{PmaxLx}[][c]{$\mathcal{P}_{\max}\cdot L^{\beta/\nu}$}
%% \psfrag{p}{$p$}
%% \psfrag{L}{$L=$}
\caption{\label{fig:PmaxLbetanuvsp-hc-2356}\textsc{hc-2,3,5,6}}
\includegraphics[width=.99\textwidth]{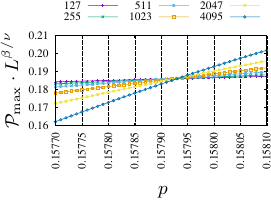}
\end{subfigure}
\hfill %% --------------------------------------------------------
\begin{subfigure}[b]{0.27\textwidth}
%% \psfrag{PmaxLx}[][c]{$\mathcal{P}_{\max}\cdot L^{\beta/\nu}$}
%% \psfrag{p}{$p$}
%% \psfrag{L}{$L=$}
\caption{\label{fig:PmaxLbetanuvsp-hc-2456}\textsc{hc-2,4,5,6}}
\includegraphics[width=.99\textwidth]{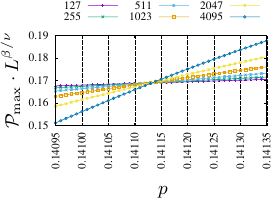}
\end{subfigure}
\hfill %% --------------------------------------------------------
\begin{subfigure}[b]{0.27\textwidth}
%% \psfrag{PmaxLx}[][c]{$\mathcal{P}_{\max}\cdot L^{\beta/\nu}$}
%% \psfrag{p}{$p$}
%% \psfrag{L}{$L=$}
\caption{\label{fig:PmaxLbetanuvsp-hc-3456}\textsc{hc-3,4,5,6}}
\includegraphics[width=.99\textwidth]{Fig-7x}
\end{subfigure}
\hfill %% --------------------------------------------------------
\begin{subfigure}[b]{0.27\textwidth}
%% \psfrag{PmaxLx}[][c]{$\mathcal{P}_{\max}\cdot L^{\beta/\nu}$}
%% \psfrag{p}{$p$}
%% \psfrag{L}{$L=$}
\caption{\label{fig:PmaxLbetanuvsp-hc-12346}\textsc{hc-1,2,3,4,6}}
\includegraphics[width=.99\textwidth]{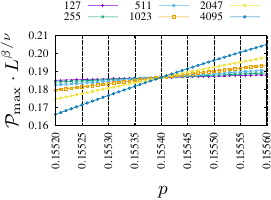}
\end{subfigure}
\hfill %% --------------------------------------------------------
\begin{subfigure}[b]{0.27\textwidth}
%% \psfrag{PmaxLx}[][c]{$\mathcal{P}_{\max}\cdot L^{\beta/\nu}$}
%% \psfrag{p}{$p$}
%% \psfrag{L}{$L=$}
\caption{\label{fig:PmaxLbetanuvsp-hc-12356}\textsc{hc-1,2,3,5,6}}
\includegraphics[width=.99\textwidth]{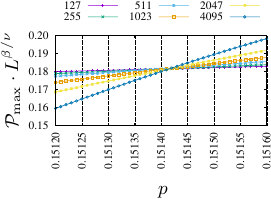}
\end{subfigure}
\hfill %% --------------------------------------------------------
\begin{subfigure}[b]{0.27\textwidth}
%% \psfrag{PmaxLx}[][c]{$\mathcal{P}_{\max}\cdot L^{\beta/\nu}$}
%% \psfrag{p}{$p$}
%% \psfrag{L}{$L=$}
\caption{\label{fig:PmaxLbetanuvsp-hc-12456}\textsc{hc-1,2,4,5,6}}
\includegraphics[width=.99\textwidth]{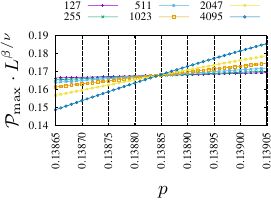}
\end{subfigure}
\hfill %% --------------------------------------------------------
\begin{subfigure}[b]{0.27\textwidth}
%% \psfrag{PmaxLx}[][c]{$\mathcal{P}_{\max}\cdot L^{\beta/\nu}$}
%% \psfrag{p}{$p$}
%% \psfrag{L}{$L=$}
\caption{\label{fig:PmaxLbetanuvsp-hc-13456}\textsc{hc-1,3,4,5,6}}
\includegraphics[width=.99\textwidth]{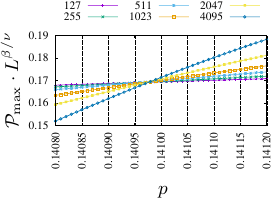}
\end{subfigure}
\hfill %% --------------------------------------------------------
\begin{subfigure}[b]{0.27\textwidth}
%% \psfrag{PmaxLx}[][c]{$\mathcal{P}_{\max}\cdot L^{\beta/\nu}$}
%% \psfrag{p}{$p$}
%% \psfrag{L}{$L=$}
\caption{\label{fig:PmaxLbetanuvsp-hc-23456}\textsc{hc-2,3,4,5,6}}
\includegraphics[width=.99\textwidth]{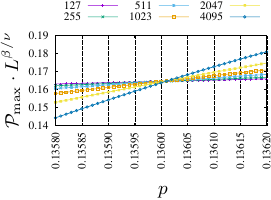}
\end{subfigure}
\hfill %% --------------------------------------------------------
\begin{subfigure}[b]{0.27\textwidth}
%% \psfrag{PmaxLx}[][c]{$\mathcal{P}_{\max}\cdot L^{\beta/\nu}$}
%% \psfrag{p}{$p$}
%% \psfrag{L}{$L=$}
\caption{\label{fig:PmaxLbetanuvsp-hc-123456}\textsc{hc-1,2,3,4,5,6}}
\includegraphics[width=.99\textwidth]{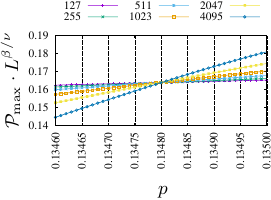}
\end{subfigure}
%% ---------------------------------------------------------------
\caption{\label{fig:PmaxLbetanuvsp-hc}$\mathcal{P}_{\max}\cdot L^{\beta/\nu}$ vs. $p$ for various complex neighborhoods on honeycomb lattice. 
Results are averaged over $R=10^5$ simulations and $\Delta p=10^{-5}$.
\subref{fig:PmaxLbetanuvsp-hc-6}      \textsc{hc-6},
\subref{fig:PmaxLbetanuvsp-hc-16}     \textsc{hc-1,6},
\subref{fig:PmaxLbetanuvsp-hc-26}     \textsc{hc-2,6},
\subref{fig:PmaxLbetanuvsp-hc-36}     \textsc{hc-3,6},
\subref{fig:PmaxLbetanuvsp-hc-46}     \textsc{hc-4,6},
\subref{fig:PmaxLbetanuvsp-hc-56}     \textsc{hc-5,6},
\subref{fig:PmaxLbetanuvsp-hc-126}    \textsc{hc-1,2,6},
\subref{fig:PmaxLbetanuvsp-hc-136}    \textsc{hc-1,3,6},
\subref{fig:PmaxLbetanuvsp-hc-146}    \textsc{hc-1,4,6},
\subref{fig:PmaxLbetanuvsp-hc-156}    \textsc{hc-1,5,6},
\subref{fig:PmaxLbetanuvsp-hc-236}    \textsc{hc-2,3,6},
\subref{fig:PmaxLbetanuvsp-hc-246}    \textsc{hc-2,4,6},
\subref{fig:PmaxLbetanuvsp-hc-256}    \textsc{hc-2,5,6},
\subref{fig:PmaxLbetanuvsp-hc-346}    \textsc{hc-3,4,6},
\subref{fig:PmaxLbetanuvsp-hc-356}    \textsc{hc-3,5,6},
\subref{fig:PmaxLbetanuvsp-hc-456}    \textsc{hc-4,5,6},
\subref{fig:PmaxLbetanuvsp-hc-1236}   \textsc{hc-1,2,3,6},
\subref{fig:PmaxLbetanuvsp-hc-1246}   \textsc{hc-1,2,4,6},
\subref{fig:PmaxLbetanuvsp-hc-1256}   \textsc{hc-1,2,5,6},
\subref{fig:PmaxLbetanuvsp-hc-1346}   \textsc{hc-1,3,4,6},
\subref{fig:PmaxLbetanuvsp-hc-1356}   \textsc{hc-1,3,5,6},
\subref{fig:PmaxLbetanuvsp-hc-1456}   \textsc{hc-1,4,5,6},
\subref{fig:PmaxLbetanuvsp-hc-2346}   \textsc{hc-2,3,4,6},
\subref{fig:PmaxLbetanuvsp-hc-2356}   \textsc{hc-2,3,5,6},
\subref{fig:PmaxLbetanuvsp-hc-2456}   \textsc{hc-2,4,5,6},
\subref{fig:PmaxLbetanuvsp-hc-3456}   \textsc{hc-3,4,5,6},
\subref{fig:PmaxLbetanuvsp-hc-12346}  \textsc{hc-1,2,3,4,6},
\subref{fig:PmaxLbetanuvsp-hc-12356}  \textsc{hc-1,2,3,5,6},
\subref{fig:PmaxLbetanuvsp-hc-12456}  \textsc{hc-1,2,4,5,6},
\subref{fig:PmaxLbetanuvsp-hc-13456}  \textsc{hc-1,3,4,5,6},
\subref{fig:PmaxLbetanuvsp-hc-23456}  \textsc{hc-2,3,4,5,6},
\subref{fig:PmaxLbetanuvsp-hc-123456} \textsc{hc-1,2,3,4,5,6}}
\end{figure*}
%% ===============================================================

%% =============================================================
\bibliography{bib/percolation,bib/basics,bib/km}
%% =============================================================

%% #############################################################
%% #############################################################
%% #############################################################
\end{document}